\UseRawInputEncoding

\pdfoutput=1

\documentclass[11pt]{article}

\usepackage{ACL2023}
\usepackage{amsmath}
\usepackage{amssymb}
\usepackage{times}
\usepackage{latexsym}
\usepackage{enumitem}
\usepackage{arydshln}
\usepackage{booktabs}
\usepackage{multirow}

\usepackage{tcolorbox} 
\tcbuselibrary{skins,vignette,raster,listings, breakable, theorems}
\usepackage{xcolor}
\usepackage{svg}
\usepackage{bm}
\usepackage{afterpage}
\usepackage{mdframed}
\tcbuselibrary{skins, breakable}

\usepackage{arydshln}
\usepackage{color}
\usepackage{multirow}
\usepackage{tabularx}
\usepackage{makecell}
\usepackage{longtable}

\usepackage{hhline}
\newcommand{\emb}[1]{\texttt{{\color{myblue}{#1}}}}
\newcommand{\newemb}[1]{\texttt{{\color{myblue2}{#1}}}}
\definecolor{myblue}{HTML}{CF553D}
\definecolor{myblue2}{HTML}{4472C4}
\usepackage{enumitem}  

\usepackage{graphicx}
\usepackage{array}

\usepackage{multirow}
\usepackage{colortbl} 
\usepackage{arydshln}
\usepackage{booktabs}

\usepackage[T1]{fontenc}

\usepackage[utf8]{inputenc}

\usepackage{microtype}

\usepackage{inconsolata}
\usepackage{graphicx}
\makeatletter
\newcommand*\bigcdot{\mathpalette\bigcdot@{.5}}
\newcommand*\bigcdot@[2]{\mathbin{\vcenter{\hbox{\scalebox{#2}{$\m@th#1\bullet$}}}}}
\makeatother

\title{MolTC: Towards Molecular Relational Modeling In Language Models}

\author{Junfeng Fang$^{1}$\footnotemark[1] \quad Shuai Zhang$^{1}$\footnotemark[1] \quad Chang Wu$^1$\quad Zhengyi Yang$^1$ \\ \textbf{Zhiyuan Liu$^2$ \quad Sihang Li$^{1}$ \quad Kun Wang$^1$ \quad Wenjie Du$^{1}$\footnotemark[2] \quad Xiang Wang$^{1}$\footnotemark[2]}\\
$^1$University of Science and Technology of China \quad 
$^2$National University of Singapore\\
\texttt{\{fjf,shuaizhang,wuchang0124,yangzhy\}@mail.ustc.edu.cn}\\ \texttt{\{acharkq,sihang0520\}@gmail.com}, \texttt{\{wk520529,duwenjie\}@mail.ustc.edu.cn}\\
\texttt{xiangwang1223@gmail.com}}

\begin{document}
\maketitle

\begin{abstract}
Molecular Relational Learning (MRL), aiming to understand interactions between molecular pairs, plays a pivotal role in advancing biochemical research.
Recently, the adoption of large language models (LLMs), known for their vast knowledge repositories and advanced logical inference capabilities, has emerged as a promising way for efficient and effective MRL.
Despite their potential, these methods predominantly rely on textual data, thus not fully harnessing the wealth of structural information inherent in molecular graphs.
Moreover, the absence of a unified framework exacerbates the issue of insufficient data exploitation, as it hinders the sharing of interaction mechanisms learned across various datasets.
To address these challenges, this work proposes a novel LLM-based multi-modal framework for \textbf{Mol}ecular in\textbf{T}eraction modeling following \textbf{C}hain-of-Thought (CoT) theory, termed \textbf{MolTC}, which effectively integrate graphical information of two molecules in pair. 
To train MolTC efficiently, we introduce a \textit{Multi-hierarchical CoT} theory to refine its training paradigm, and conduct a comprehensive \textbf{Molecular Interactive Instructions} dataset for the development of biochemical LLMs involving MRL. 
Our experiments, conducted across twelve datasets involving over 4,000,000 molecular pairs, exhibit the superiority of our method over current GNN- and LLM-based baselines. 
Our code is available at 
\href{https://github.com/MangoKiller/MolTC}{https://github.com/MangoKiller/MolTC}.
\end{abstract}
\section{Introduction}
\renewcommand{\thefootnote}{\fnsymbol{footnote}}
\footnotetext[1]{Equal contribution.}
\footnotetext[2]{Corresponding author. Xiang Wang is also affiliated with Institute of Dataspace, Hefei Comprehensive National Science Center.}
\renewcommand{\thefootnote}{\arabic{footnote}}

Molecular Relational Learning (MRL) \citep{CGIB},
aiming to understand interactions between molecular \textit{pairs}, has gained significant interest due to its wide range of applications \cite{drug_drug_3}. For example, Drug-Drug Interactions (DDIs) are
critical in pharmacology and drug development \cite{drug_drug_1},
while solute-solvent interactions (SSIs) are fundamental in solution chemistry and the design of chemical processes \cite{SSI_1,SSI_2}. 
However, the exhaustive experimental validation of these interactions is notoriously time-consuming and costly. In response, adopting large language models (LLMs) \cite{LLM_1,Galactica}, known for their vast knowledge repositories and advanced logical inference capabilities,  has emerged as an efficient and effective alternative for MRL \cite{LLM4MI1,LLM4MI2}.

\begin{figure*}[t]
    \centering
    \includegraphics[width=1.02\linewidth]{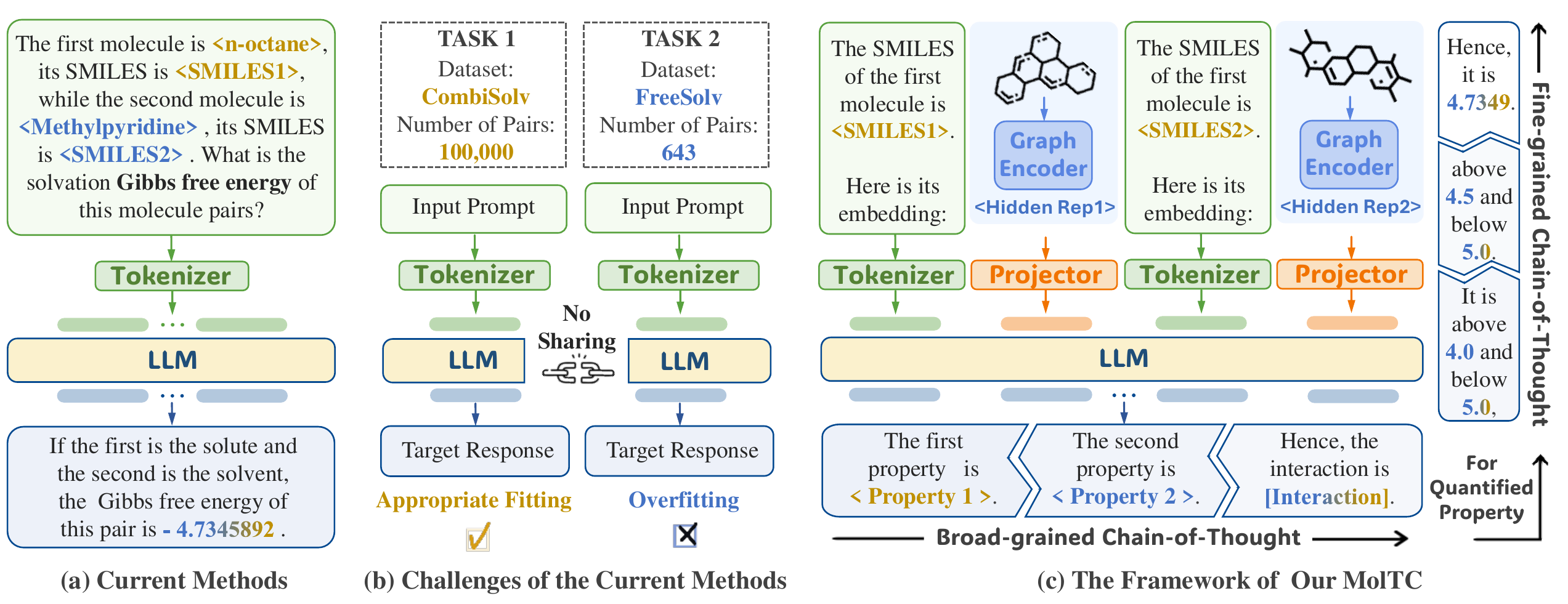}
    \caption{Comparison between the current methods leveraging LLMs to model molecule interactions and our MolTC. (a) The prevailing paradigm of current methods. (b) The challenge of applying the current paradigm to the tasks involving datasets with a small number of samples. (c) The framework of our  proposed MolTC, which is enhanced by the principle of CoT. Best viewed in color.}
    \label{fig:intro}
\end{figure*}

Despite their promise, 
a primary concern of current LLM-based paradigm is the \textit{insufficient         data exploitation}.
Specifically, they predominantly rely on the textual data such as SMILES (Simplified Molecular Input Line Entry System) and property descriptions, 
thus not fully harnessing the wealth of structural information inherent in molecular graphs \cite{LLM4MI4},
as indicated in Figure \ref{fig:intro} (a). Current studies have indicated that it is challenging for LLMs to fully understand the complex graphs based solely on textual data, hence, it's crucial to explicitly model these structures given their significance in MRL  \cite{LLM4MI1}.

Compounding this concern is the absence of a unified framework for LLM-based MRL \cite{LLM4MI6,LLM4MI7}. Concretely, this absence impedes the sharing and integration of interaction mechanisms learned across various datasets, leading to a fragmentation in collective insights. Especially, it poses a catastrophic challenge for tasks with a limited number of labeled pairs \cite{SSI_2}, where LLMs often struggle with due to the high risk of overfitting, as illustrated in Figure \ref{fig:intro} (b). Worse still, such limited datasets are prevalent in MRL since the experimental acquisition is often constrained by high costs \cite{CGIB}.

To overcome these limitations, in this work, we propose \textbf{MolTC}, a unified multi-modal framework for \textbf{Mol}ecular in\textbf{T}eraction modeling following the \textbf{Chain}-of-thought theory \cite{CoT}.
As depicted in Figure \ref{fig:intro} (c), MolTC employs the Graph Neural Networks (GNNs) \cite{GCN}, known for their proficiency in graph modeling, to explicitly gather graphical information of molecular pairs, and integrates them into the input space of LLMs by two meticulously crafted projectors. In response to empirical findings that LLMs may confuse two input molecules in pair, MolTC incorporates the molecules' SMILES information to reinforce the concept of molecular order. 

More importantly, a two-pronged approach is developed to train  MolTC efficiently:

\vspace{3pt}
\noindent\textbf{(1) Training Paradigm Refinement:} As shown in Figure \ref{fig:intro} (c), we introduce a \textit{Multi-hierarchical CoT} theory to guide the training paradigm of MolTC. Concretely, the broad-grained CoT guides the pre-training stage to identify individual molecular properties before predicting interactions, ensuring an acute awareness of each molecule's unique attribute. For quantitative interaction tasks, which are challenging for LLMs, a fine-grained CoT enables the fine-tuning stage to initially predict a range, and then progressively refining it to a precise value.

\vspace{3pt}
\noindent\textbf{(2) Dataset Foundation Construction:} 
In sight of the absence of a comprehensive MRL
datasets for biochemical LLMs,  we construct a \textbf{Mol}ecular in\textbf{T}eractive instructions dataset, termed \textbf{MoT-instruction}. Specifically, we first conduct twelve well-established MRL datasets across various domain, and source their detailed molecular properties from authoritative biochemical databases. Based on this, we meticulously compile these properties and empirically determine their optimal instructions.
These process ensures that MoT-instructions can not only enhance the performance of our MolTC, but also  contribute to the development of other biochemical LLMs involving MRL. 
    
Our contributions can be summarized as follows:
\begin{itemize}[leftmargin=*]
    \setlength{\itemsep}{0pt}
    \setlength{\parsep}{0pt}
    \setlength{\parskip}{0pt}
    \item We identify the issue of insufficient data exploitation in current LLM-based MRL, and take the first attempt to develop a unified multi-modal framework for LLM-based MRL, named MolTC.
    \item We introduce the multi-hierarchical CoT theory to enhance the MolTC's training process, especially for quantitative interaction tasks. 
    \item We construct MoT-instructions, the first comprehensive instruction dataset in MRL domain, to enhance the development of biochemical LLMs involving MRL.
    \item Our experiments, across over 4,000,000 molecular pairs in various domains such as DDI and SSI, demonstrate the superiority of our method over current GNN and LLM-based baselines.
\end{itemize}

\section{Methodology}
In this section, we detail our MolTC, which harnesses the power of LLMs for comprehending molecular interactions. We begin with the introduction of model framework in Section \ref{sec:model_arch}. Taking a step further, the training paradigm guided by the principle of Multi-hierarchical CoT is outlined in Section \ref{sec:training}. Moreover, the dynamic parameter sharing strategy tailored for MolTC and our developed datasets, MoT-instructions, are elaborated in Section \ref{sec:parameter} and \ref{sec:instruction}, respectively.

\subsection{Framework of MolTC} \label{sec:model_arch}
Here we introduce four key components of 
MolTC’s framework: Graph Encoder, Representation Projector, SIMLES Injector, and the backbone LLM. The specific instantiation details can be found in the experimental section and the appendix.

\vspace{5pt}
\noindent\textbf{Graph Encoder.}
The first step of extracting interactions is to precisely encode the molecular graphs. In sight of this, we utilize two GNN-based encoders to capture the embedding of the given molecular pairs, leveraging the GNN's robust capability in aggregating structural information.  More formally, let $\mathcal{G}_a =\{\mathcal{V}_a,\mathcal{E}_a\}$ and $\mathcal{G}_b=\{\mathcal{V}_b,\mathcal{E}_b\}$ denote the input pair, where $\mathcal{V},\mathcal{E}$ represent atomic nodes and the chemical bonds, respectively. The two graph encoders $f_{\texttt{enc1}}$ and $f_{\texttt{enc2}}$ perform aggregating to obtain the atomic embedding:
\begin{equation}
\begin{aligned}
    &\mathbf{H}_a=[\boldsymbol{h}_a^1, \boldsymbol{h}_a^2, \ldots, \boldsymbol{h}_a^{|\mathcal{V}_a|}]=f_{\texttt{enc1}}(\mathcal{G}_a),\\
    &\mathbf{H}_b\,=[\boldsymbol{h}_b^1, \boldsymbol{h}_b^2, \ldots, \boldsymbol{h}_b^{|\mathcal{V}_b|}]=f_{\texttt{enc2}}(\mathcal{G}_b),
\end{aligned}
\end{equation}
where $\boldsymbol{h}_a^i$ and $\boldsymbol{h}_b^i$ denote to the embedding of the $i$-th atom in molecule $\mathcal{G}_a$ and $\mathcal{G}_b$; $\mathcal{V}_a$ and $\mathcal{V}_b$ represent the number of nodes.

\vspace{5pt}
\noindent\textbf{Representation Projector.}
After acquiring molecular pair representations $\mathbf{H}_a$ and $\mathbf{H}_b$, the next step is to map them into the backbone LLM's hidden space using Projectors $f_{\texttt{pro1}}$ and $f_{\texttt{pro2}}$. These projectors serve as pivotal connectors, translating $\mathbf{H}_a$ and $\mathbf{H}_b$ into LLM-comprehensible encodings $\mathbf{M}_a$ and $\mathbf{M}_b$. 
Drawing inspiration from the state-of-the-art vision-language models, we instantiate $f_{\texttt{pro1}}$ and $f_{\texttt{pro2}}$ by Querying Transformers (Q-Formers) \cite{QFormer,qFormer2}. 
More formally,
\begin{equation}
\begin{aligned}
    &\mathbf{M}_a=[\boldsymbol{m}_a^1, \boldsymbol{m}_a^2, \ldots, \boldsymbol{m}_a^{q}]=f_{\texttt{pro1}}(\mathbf{H}_a),\\
    &\mathbf{M}_b\,=[\boldsymbol{m}_b^1, \boldsymbol{m}_b^2, \ldots, \boldsymbol{m}_b^{q}]=f_{\texttt{pro2}}(\mathbf{H}_b),
\end{aligned}
\end{equation}
where $q$ denotes the number of learnable query tokens of Q-Former's transformer. 

In detail, our Projectors, based on the BERT architecture, incorporate an additional cross-attention module positioned between the self-attention and feed-forward modules. 
This instantiation offers two key benefits. Firstly, it supports seamless integration with conventional BERT-based text encoders, allowing $f_{\texttt{pro1}}$ and $f_{\texttt{pro2}}$ pre-training with extensive molecular graph-text pairs. Secondly, it maintains compatibility with various input dimensions $d$, and allows adjustments in the size of learnable query tokens to align with the LLM's token embedding size. These advantages lay a solid foundation for the thorough interaction of two molecules during the LLM's inference process. Future work will also explore more projector designs, such as streamlining it through specially tailored MLPs \cite{yangzhengyi}.

\vspace{5pt}
\noindent \textbf{SMILES Tokenization.}
When directly analyzing the representations $\mathbf{M}_a$ and $\mathbf{M}_b$ with LLMs, our experiments suggest a potential confusion by LLMs in distinguishing the properties of each molecule in a pair. This observation naturally inspires us to integrate textual information of the molecules to strengthen the concept of their sequential order. Here MolTC employs SMILES due to its ubiquity and specificity. Additionally, SMILES serves as a conduit, linking the task-specific prompts with the corresponding biochemical knowledge stored within the LLM. Therefore, we directly input the SMILES of both molecules into the backbone LLM, utilizing the inherent encoder to acquire their tokens $\mathbf{S}_a$ and $\mathbf{S}_b$.

\vspace{5pt}
\noindent \textbf{Backbone LLM.}
MolTC leverages Galactica, a decoder-only transformer built on the OPT framework, as its base LLM. 
Pretrained on an extensive collection of scientific literature, Galactica demonstrates exceptional proficiency in biochemistry knowledge. This expertise, particularly in parsing molecular sequences such as SMILES and SELFIES strings, enables Galactica to adeptly capture the properties crucial for molecular interactions. Specifically, the goal of MolTC is to harness Galactica's advanced inferential skills to interpret the contextual interactions between two molecular sets of token collections, $\{\mathbf{M}_a, \mathbf{S}_a\}$ and $\{\mathbf{M}_b, \mathbf{S}_b\}$. More formally, we denote an integrated prompt sequence as follows:
\begin{equation}
\begin{aligned}
    \mathbf{X}=\{\mathbf{P},\mathbf{M}_a, &\mathbf{S}_a,\mathbf{M}_b,\mathbf{S}_b\} = \left[\boldsymbol{x}_1, \boldsymbol{x}_2, ..., \boldsymbol{x}_l\right] \\
    &\,\,\,\,\text{ s.t. } \mathbf{P}\sim\mathcal{P}_{\mathbf{r}},
\end{aligned}
\end{equation}
where $l$ is the integrated input length, $\mathbf{P}$ denotes the task-specific prompt, and $\mathcal{P}_{\mathbf{r}}$ represents a collection of various manually designed prompts, each tailored for the molecular interaction task $\mathbf{r}$.
The generation process adopts a causal mask to generate a response encapsulating key interactive properties with length $T$:
\begin{equation}
\begin{aligned}
    \mathbf{\hat{X}} = [\boldsymbol{\hat{x}}_1, \boldsymbol{\hat{x}}_2, ..., \boldsymbol{\hat{x}}_T].
\end{aligned}
\end{equation}
Utilizing Galactica's autoregressive framework, the training objective involves regressing the target response based on the input prompt $\mathbf{X}$. Specifically, the output for $i$-th token $\boldsymbol{\hat{x}}_i$, is computed based on its preceding tokens as follows for $t\in (1,T)$: 
\begin{equation}
p\left(\mathbf{\hat{X}}_{[1: t]}|\mathbf{X}\right)=\prod_{i=1}^t p\left(\boldsymbol{\hat{x}}_i|\mathbf{X}, \mathbf{\hat{X}}_{[1: i-1]}\right).
\end{equation}

\subsection{Training Paradigm of MolTC} \label{sec:training}
\begin{figure}[t]
    \centering
    \includegraphics[width=1.03\linewidth]{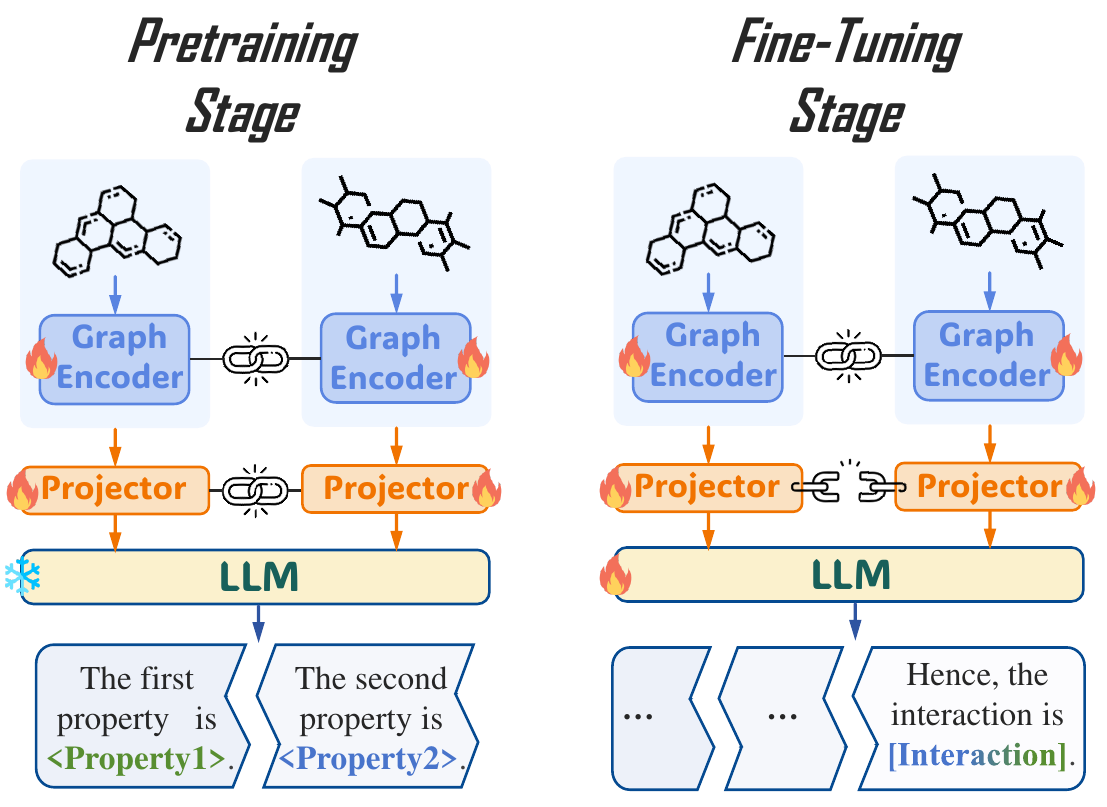}
    \caption{The training process of our MolTC. The flame symbol
 denotes the parameter update, the snowflake symbol
 indicates the parameter freezing, and the chain symbol
  depicts the parameter sharing between two modules. Best viewed in color.}
    \label{fig:intro2}
\end{figure}

In this part, we elaborate the training paradigm of MolTC, including pretraining and fine-tuning processes, which is guided by the principle of Multi-hierarchical CoT, as shown in Figure \ref{fig:intro2}.

\subsubsection{Broad-grained CoT Guided Pretraining}

Given the challenge of directly understanding complex interactions between two input molecules in pair,
the broad-grained CoT guides MolTC to initially identify individual molecular properties. By thoroughly understanding each molecule's characteristics, MolTC establishes a solid foundation for accurately predicting their interactions. 
Specifically, in the pretraining stage, the prompt is uniformly designed as follows:
\begin{tcolorbox}[boxrule=0.5pt,left=0pt,right=0pt,top=2.5pt,bottom=2.5pt,title={Prompt for Pretraining Stage}]
    \centering
    \begin{tabular}{cl}
        \makecell[c]{Input \\ Prompt} 
         &
         \makecell*[{{p{5.1cm}}}]{\emb{<SMILES1>}, \emb{<GraEmb1>}, {the front is the first molecule, followed by the second molecule:}
        \newemb{<SMILES2>}. \newemb{<GraEmb2>}. {Please provide the biochemical properties of the two molecules one by one.}}\\
        \hline
        \makecell[c]{Target \\ Response} 
         & \makecell*[{{p{5.1cm}}}]{The properties of the first molecule are \emb{[Property1]}, and the properties of the second molecule are \newemb{[Property2]}.
}
    \end{tabular}
\end{tcolorbox}

This prompt design enable MolTC to delineate key properties of two molecules sequentially. Based on it, MolTC utilize the generation loss of the backbone LLM to train Graph Encoders, $f_{\texttt{enc1}}$ and $f_{\texttt{enc2}}$, as well as the Representation Projectors, $f_{\texttt{pro1}}$ and $f_{\texttt{pro2}}$. Notably, during this phase, the backbone LLM remains frozen. 

\vspace{3pt}
\noindent \textbf{Dataset Construction for Pretraining.} 
To ensure backbone LLM can understand the individual characteristics of each molecule, it is pivotal to prepare a comprehensive dataset comprising molecule pairs and their corresponding biochemical properties. To this end, 
(1) we first conduct an extensive survey of various authoritative biochemical database such as PubChem\footnote{ \href{https://pubchem.ncbi.nlm.nih.gov}{https://pubchem.ncbi.nlm.nih.gov}} and Drugbank \cite{PubChem}, and collect a large amount of molecule-textual properties pairs; (2) then, recognizing the variability in annotation quality within this dataset, we augment and enrich molecular descriptions that were less extensively annotated; (3) subsequently, to simulate diverse molecular interactions, we generated molecular pairs by randomly grouping two distinct molecules from the above database. This random pairing facilitates a broad spectrum of molecular combinations, exposing the pretraining stage to diverse interaction scenarios, thus naturally enhancing the generalizability of our MolTC.

\subsubsection{Fine-grained CoT Guided Fine-tuning}
During the fine-tuning phase, MolTC is trained to enable the backbone LLM to generate interaction properties based on the properties of individual molecules it initially identifies. To this end, prompts in the fine-tuning stage should be crafted for specific downstream task. For example, in DDI tasks, we construct the following prompt:

\begin{tcolorbox}[boxrule=0.5pt,left=0pt,right=0pt,top=2.5pt,bottom=2.5pt,title={Prompt for DDI Tasks (Fine-tuning)}]
    \centering
    \begin{tabular}{cl}
        \makecell[c]{Input \\ Prompt} 
         &
         \makecell*[{{p{5.1cm}}}]{\emb{<SMILES1>}, \emb{<GraEmb1>}, {the front is the first molecule, followed by the second molecule:}
        \newemb{<SMILES2>}. \newemb{<GraEmb2>}. {What are the side effects of these two drugs?}}\\
        \hline
        \makecell[c]{Target \\ Response} 
         & \makecell*[{{p{5.1cm}}}]{The property of the first molecule is \emb{[Property1]}, while the property of the second molecule is \newemb{[Property2]}. Hence, the first drug molecule may increase the photosensitizing activities of the second drug molecule.}
    \end{tabular}
\end{tcolorbox}

Despite the effectiveness of this prompt design, LLMs face notable challenges in quantitative analysis, especially in complex molecular interaction contexts such as SSI and chromophore-solvent interaction (CSI). Our experiments in Section \ref{sec:exp} highlight this difficulty, demonstrating that LLMs tend to exhibit indecision regarding the quantitative values in their outputs. 
To address this, a fine-grained CoT concept is introduced to refine the training paradigm. Specifically, the backbone LLM is guided to initially suggest a range for the target numerical value, then progressively refining it to a precise value. Take a meticulously prompt for SSI tasks as an example: 

\begin{tcolorbox}[boxrule=0.5pt,left=0pt,right=0pt,top=2.5pt,bottom=2.5pt,title={Prompt for SSI Tasks (Fine-tuning)}]
    \centering
    \begin{tabular}{cl}
        \makecell[c]{Input \\ Prompt} 
         &
         \makecell*[{{p{5.1cm}}}]{\emb{<SMILES1>}, \emb{<GraEmb1>}, {the front is the first molecule, followed by the second molecule:}
        \newemb{<SMILES2>}. \newemb{<GraEmb2>}. {What is the solvation Gibbs free energy of this pair of molecules?}}\\
        \hline
        \makecell[c]{Target \\ Response} 
         & \makecell*[{{p{5.1cm}}}]{The property of the first molecule is \emb{[Property1]}, while the property of the second molecule is \newemb{[Property2]}. Hence, the solvation Gibbs free energy of these two molecules is above \texttt{3.0} and below \texttt{3.5}, so the accurate value is \texttt{3.24791}.
}
    \end{tabular}
\end{tcolorbox}

This step-wise refinement process fosters a more accurate and reliable resolution of numerically-intensive challenges.
Based on these prompts, in the fine-tuning stage, the  parameters in backbone LLM are updated through Low-Rank Adaptation (LoRA) \cite{LoRA} strategy, known for its efficiency in tailoring the LLM to the requirements of downstream tasks and minimal memory demands in storing gradients. Meanwhile, to ensure that other modules are optimally adjusted to suit the specifics of the downstream tasks, Graph Encoders $f_{\texttt{enc1}}$ and $f_{\texttt{enc2}}$, as well as Representation Projectors $f_{\texttt{pro1}}$ and $f_{\texttt{pro2}}$ are trained following the generation loss of the backbone LLM.

\subsection{Dynamic Parameter Sharing Strategy} \label{sec:parameter}
To implement the above training paradigm effectively, we introduce a novel parameter-sharing strategy, inspired by key biochemical insights:

\vspace{3pt}
\noindent (1) The Importance of \textbf{Role-Playing}: A molecule's role in an interaction crucially influences the outcome. For example, in SSI scenario like the water-ethanol pair, utilizing water and ethanol as solvents, respectively, yields different energy releases \cite{SSI-P1}. Sometimes, a reversal of roles can even result in the absence of interaction.

\vspace{3pt}
\noindent (2) The Importance of \textbf{Input Order}: In certain molecular pairs, the sequence of introducing molecules significantly impacts the interactions. For instance, the order of drug introduction can lead to varying therapeutic effects.

\vspace{3pt}
\noindent (3) The Importance of \textbf{Role and Order-Specific} Feature Extraction: The role and input order of molecules determine the relevance of their structural features. For example, a chemical group in a solute-solvent pair may be crucial for the release of Gibbs free energy when in the solute, but less so in the solvent \cite{SSI-P1,SSI-P2}.

These insights inspire MolTC to adaptively prioritize distinct key information, creating unique tokens for the same molecule based on its role and order. To enable this nuanced learning while also capitalizing on the shared aspects of molecular learning, we introduce the following parameter-sharing strategy, as shown in Figure \ref{fig:intro2}:

\vspace{3pt}
\noindent (1) The GNN-based \textbf{Encoders} $f_{\texttt{enc1}}$ and $f_{\texttt{enc2}}$, which focus on extracting molecular graph structures, share parameters during both pretraining and fine-tuning stages to enhance learning efficiency.

\vspace{3pt}
\noindent (2) The Qformer-based \textbf{Projectors} $f_{\texttt{pro1}}$ and $f_{\texttt{pro2}}$, tasked with aligning molecular structures to semantic information, share parameters during pretraining stage to promote generalization and robustness. However, in the fine-tuning stage, we cease sharing to allow customized semantic mappings tailored to the varying roles and orders.

In summary, this strategy is tailored to balance the need for role and order-based distinctively learning with the efficiency gained from commonalities across molecular pairs.

\subsection{Construction of MoT-instructions}  \label{sec:instruction}
Given the absence of a comprehensive instruction datasets tailored for LLM-based MRL, we aim to develop a molecular interactive instructions dataset, termed MoT-instructions. This dataset is designed to fulfill several key criteria: (1) it should include extensive molecular pairs capable of interaction, covering a broad spectrum of domains, (2) it should detail important biochemical properties of each molecule within these pairs, and (3) it should elaborate the resultant properties from molecular interactions. Specifically, MoT-instructions are constructed through a three-step process as follows.

\vspace{3pt}
\noindent(1) We begin by aggregating twelve representative molecular interaction datasets across various widely recognized biochemical tasks, such as DDI, SSI, and CSI. Following this, we engage in a systematic search for textual descriptions of the biochemical properties of each molecule involved in these interactions. Specifically, we source this information from authoritative biochemical databases such as DrugBank and PubChem.

\vspace{3pt}
\noindent(2) The next critical step is the \textbf{experimental determination of the optimal instructions}. Specifically, for all molecular pairs in step (1), we first deconstruct the lengthy molecular properties into a series of questions and answers, a format more comprehensible to LLMs \cite{Galactica}. The granularity of this deconstruction is decided based on the performance of our MolTC. For more challenging quantitative tasks, instructions guided by fine-grained CoT are required to provide a numerical range before specifying a concrete value. Given the vast number of possible correct ranges, exhaustive testing is impractical. Therefore, we initially determine the   optimal range for a small subset of datasets using a grid search, guided by the predictive performance of MolTC. Subsequently, we derive statistics, such as mean and standard deviation, from these datasets to establish a relationship between statistics and optimal ranges. Finally, for other datasets, we determine their optimal range based on this established rule.

\vspace{3pt}
\noindent(3) The final step in our dataset construction involved filtering out pairs that lacked sufficient information on molecule properties or interaction data. Specifically, partial properties of a molecular pair are often missing in some datasets. To maximize the utilization of information from these datasets, we consider extracting each property within them as a separate dataset. This approach allows us to naturally omit missing values without wasting other information present in the molecular pair.

\section{Experiment} \label{sec:exp}
In this section, we aim to answer the following research questions:
\vspace{-4pt}
\begin{itemize}[leftmargin=*]
    \setlength{\itemsep}{0pt}
    \setlength{\parsep}{0pt}
    \setlength{\parskip}{0pt}
    \item \textbf{RQ1:} Is MolTC capable of generating the interactive property, involving the \textit{qualitative}   knowledge, of the given molecular pair?
    \item \textbf{RQ2:} Does MolTC have the ability to generate the interactive property,  involving the \textit{quantitative} property, for a given molecular pair?
    \item \textbf{RQ3:}  What is the impact of the proposed strategies, such as the CoT enhancement strategy and SMILES injection strategy, on the inference process of our MolTC?
\end{itemize}

\begin{table*}[ht]\footnotesize
\centering
\caption{Comparative performance of various methods in qualitative interactive tasks. The best-performing methods are highlighted with a gray background, while the second-best methods are underscored for emphasis.}
\vspace{-0.6em}
\setlength{\tabcolsep}{2pt}
\def\arraystretch{1.07}
\label{tab:performance_comparison}
\begin{tabular}{cc|cc|cc|cc|cc}
\toprule
\multicolumn{2}{c|}{\multirow{2}{*}{ \thead{Baseline   Model}}}  & \multicolumn{2}{c|}{ Drugbank} & \multicolumn{2}{c|}{ ZhangDDI} & \multicolumn{2}{c|}{ ChChMiner} & \multicolumn{2}{c}{ DeepDDI} \\ 
                     &     &  Accuracy           &  AUC-ROC         &  Accuracy           &  AUC-ROC         &  Accuracy            &  AUC-ROC          &  Accuracy           &  AUC-ROC        \\ \midrule  \midrule

\multicolumn{1}{c}{\multirow{5}{*}{\thead{GNN \\ Based}}} & GoGNN      & \multicolumn{1}{c}{$84.78_{\pm0.57}$} & $91.63_{\pm0.66}$ & \multicolumn{1}{c}{$84.10_{\pm0.46}$} & $92.35_{\pm0.48}$ & \multicolumn{1}{c}{$91.17_{\pm0.46}$} & $96.64_{\pm0.40}$ & \multicolumn{1}{c}{$93.54_{\pm0.35}$} & $92.71_{\pm0.27}$ \\ 

\multicolumn{1}{c}{}      & SSI-DDI    & \multicolumn{1}{c}{$94.12_{\pm0.33}$} & $98.38_{\pm0.31}$ & \multicolumn{1}{c}{$86.97_{\pm0.37}$} & $93.76_{\pm0.34}$ & \multicolumn{1}{c}{$93.26_{\pm0.31}$} & $97.81_{\pm0.22}$ & \multicolumn{1}{c}{$95.27_{\pm0.25}$} & $98.42_{\pm0.31}$ \\

\multicolumn{1}{c}{}                           & DSN-DDI    & \multicolumn{1}{c}{{$\underline{94.93}_{\pm0.14}$}} & {$\underline{99.01}_{\pm0.12}$} & \multicolumn{1}{c}{$87.65_{\pm0.13}$} & $94.63_{\pm0.18}$ & \multicolumn{1}{c}{$84.30_{\pm0.17}$} & $94.25_{\pm0.26}$ & \multicolumn{1}{c}{$95.64_{\pm0.18}$} & $98.01_{\pm0.16}$ \\

\multicolumn{1}{c}{}                           & CMRL       & \multicolumn{1}{c}{$94.83_{\pm0.12}$} & $98.76_{\pm0.10}$ & \multicolumn{1}{c}{{$\underline{87.78}_{\pm0.36}$}} & {$\underline{94.68}_{\pm0.23}$} & \multicolumn{1}{c}{$94.23_{\pm0.26}$} & $98.37_{\pm0.12}$ & \multicolumn{1}{c}{{$\underline{96.37}_{\pm0.34}$}} & {$\underline{98.98}_{\pm0.31}$} \\ 

\multicolumn{1}{c}{}                           & CGIB       & \multicolumn{1}{c}{$94.68_{\pm0.34}$} & $98.60_{\pm0.25}$ & \multicolumn{1}{c}{$87.32_{\pm0.71}$} & $94.18_{\pm0.60}$ & \multicolumn{1}{c}{{$\underline{94.25}_{\pm0.39}$}} & {$\underline{98.45}_{\pm0.31}$} & \multicolumn{1}{c}{$96.23_{\pm0.52}$} & $98.45_{\pm0.64}$ \\
  \midrule

\multicolumn{1}{c}{\multirow{3}{*}{\thead{ML \\ Based}}}  & DeepDDI    & \multicolumn{1}{c}{$93.15_{\pm0.25}$} & $98.06_{\pm0.54}$ & \multicolumn{1}{c}{$83.35_{\pm0.49}$} & $91.13_{\pm0.58}$ & \multicolumn{1}{c}{$90.34_{\pm0.62}$} & $95.73_{\pm0.37}$ & \multicolumn{1}{c}{$92.39_{\pm0.38}$} & $98.11_{\pm0.42}$ \\

\multicolumn{1}{c}{}                            & MHCADDI    & \multicolumn{1}{c}{$78.50_{\pm0.80}$} & $86.33_{\pm0.35}$ & \multicolumn{1}{c}{$77.86_{\pm0.59}$} & $86.94_{\pm0.68}$ & \multicolumn{1}{c}{$84.26_{\pm0.54}$} & $89.33_{\pm0.82}$ & \multicolumn{1}{c}{$87.01_{\pm0.77}$} & $88.64_{\pm0.83}$ \\

\multicolumn{1}{c}{}                           & MDF-SA-DDI & \multicolumn{1}{c}{$93.86_{\pm0.31}$} & $97.65_{\pm0.29}$ & \multicolumn{1}{c}{$86.89_{\pm0.25}$} & $94.03_{\pm0.23}$ & \multicolumn{1}{c}{$93.64_{\pm0.20}$} & $98.10_{\pm0.19}$ & \multicolumn{1}{c}{$95.12_{\pm0.30}$} & $97.84_{\pm0.36}$ \\  \midrule

\multicolumn{1}{c}{\multirow{4}{*}{\thead{LLM \\ Based}}} & Galactica  & \multicolumn{1}{c}{$79.16_{\pm0.35}$} & $86.23_{\pm0.33}$ & \multicolumn{1}{c}{$67.20_{\pm0.46}$} & $78.74_{\pm0.58}$ & \multicolumn{1}{c}{$74.61_{\pm0.44}$} & $83.51_{\pm0.63}$ & \multicolumn{1}{c}{$71.50_{\pm0.41}$} & $79.07_{\pm0.41}$ \\ 

\multicolumn{1}{c}{}                           & Chem T5    & \multicolumn{1}{c}{$85.83_{\pm0.31}$} & $91.97_{\pm0.38}$ & \multicolumn{1}{c}{$72.34_{\pm0.42}$} & $89.31_{\pm0.30}$ & \multicolumn{1}{c}{$80.79_{\pm0.52}$} & $85.65_{\pm0.46}$ & \multicolumn{1}{c}{$75.58_{\pm0.66}$} & $84.42_{\pm0.43}$ \\ 

\multicolumn{1}{c}{}                           & MolCA      & \multicolumn{1}{c}{$87.95_{\pm0.52}$} & $94.00_{\pm0.37}$ & \multicolumn{1}{c}{$68.21_{\pm0.59}$} & $88.53_{\pm0.62}$ & \multicolumn{1}{c}{$90.15_{\pm0.43}$} & $92.92_{\pm0.60}$ & \multicolumn{1}{c}{$82.95_{\pm0.58}$} & $88.52_{\pm0.77}$ \\ 

\multicolumn{1}{c}{}                           & MolT5      & \multicolumn{1}{c}{$89.49_{\pm0.47}$} & $93.08_{\pm0.26}$ & \multicolumn{1}{c}{$76.46_{\pm0.30}$} & $89.06_{\pm0.33}$ & \multicolumn{1}{c}{$84.70_{\pm0.25}$} & $91.18_{\pm0.32}$ & \multicolumn{1}{c}{$86.82_{\pm0.46}$} & $90.08_{\pm0.57}$ \\ \midrule

\multicolumn{2}{c|}{ MolTC (Ours)}     & \multicolumn{1}{c}{\cellcolor{gray!30}${95.98}_{\pm0.15}$} & \cellcolor{gray!30}${99.12}_{\pm0.31}$ & \multicolumn{1}{c}{\cellcolor{gray!30}${89.40}_{\pm0.12}$} & \cellcolor{gray!30}${95.48}_{\pm0.18}$ & \multicolumn{1}{c}{\cellcolor{gray!30}${95.59}_{\pm0.20}$} & \cellcolor{gray!30}${98.66}_{\pm0.09}$ & \multicolumn{1}{c}{\cellcolor{gray!30}${96.70}_{\pm0.26}$} & \cellcolor{gray!30}${99.05}_{\pm0.32}$ \\ \bottomrule
\end{tabular}
\end{table*}

\begin{table*}[ht]\footnotesize
\centering
\caption{Comparative performance of various methods in quantitative interactive tasks. The best-performing methods are highlighted with a gray background, while the second-best methods are underscored for emphasis.}
\vspace{-0.6em}
\setlength{\tabcolsep}{1.4pt}
\def\arraystretch{1.05}
\label{tab:performance_comparison2}
\begin{tabular}{cc|cc|cc|cc|cc}
\toprule
\multicolumn{2}{c|}{\multirow{2}{*}{\thead{Baseline   Model}}}  & \multicolumn{2}{c|}{FreeSolv} & \multicolumn{2}{c|}{Abraham } & \multicolumn{2}{c|}{CompSol } & \multicolumn{2}{c}{CombiSolv } \\ 
                     &     &MAE           & RMSE         &MAE           & RMSE         &MAE            & RMSE          &MAE           & RMSE        \\ \midrule  \midrule

\multicolumn{1}{c}{\multirow{4}{*}{\thead{GNN \\ Based}}} & CIGIN      & \multicolumn{1}{c}{$0.589_{\pm0.053}$} & $0.931_{\pm0.066}$ & \multicolumn{1}{c}{$0.314_{\pm0.004}$} & $0.607_{\pm0.011}$ & \multicolumn{1}{c}{$0.197_{\pm0.003}$} & $0.349_{\pm0.005}$ & \multicolumn{1}{c}{$0.288_{\pm0.005}$} & $0.664_{\pm0.012}$ \\

\multicolumn{1}{c}{}    & D-MPNN    & \multicolumn{1}{c}{$0.702_{\pm0.014}$} & $1.231_{\pm0.029}$ & \multicolumn{1}{c}{$0.484_{\pm0.012}$} & $0.705_{\pm0.025}$ & \multicolumn{1}{c}{$0.205_{\pm0.006}$} & $0.373_{\pm0.007}$ & \multicolumn{1}{c}{$0.482_{\pm0.013}$} & $0.895_{\pm0.055}$ \\ 

\multicolumn{1}{c}{}      & GEM    & \multicolumn{1}{c}{$0.598_{\pm0.018}$} & $1.188_{\pm0.049}$ & \multicolumn{1}{c}{$\underline{0.254}_{\pm0.004}$} & $0.531_{\pm0.005}$ & \multicolumn{1}{c}{$0.203_{\pm0.006}$} & $0.337_{\pm0.007}$ & \multicolumn{1}{c}{$0.290_{\pm0.009}$} & $0.783_{\pm0.020}$ \\

\multicolumn{1}{c}{}                           & CGIB       & \multicolumn{1}{c}{$\underline{0.541}_{\pm0.009}$} & $\underline{0.917}_{\pm0.055}$ & \multicolumn{1}{c}{$0.258_{\pm0.008}$} & $\underline{0.530}_{\pm0.009}$ & \multicolumn{1}{c}{{$\underline{0.178}_{\pm0.004}$}} & {$\underline{0.301}_{\pm0.003}$} & \multicolumn{1}{c}{$\underline{0.230}_{\pm0.004}$} & $\underline{0.394}_{\pm0.009}$ \\ \midrule

\multicolumn{1}{c}{\multirow{4}{*}{\thead{ML \\ Based}}}                    & GOVER    & \multicolumn{1}{c}{$0.636_{\pm0.026}$} & $1.074_{\pm0.049}$ & \multicolumn{1}{c}{$0.347_{\pm0.005}$} & $0.625_{\pm0.016}$ & \multicolumn{1}{c}{$0.184_{\pm0.005}$} & $0.371_{\pm0.014}$ & \multicolumn{1}{c}{$0.412_{\pm0.016}$} & $0.728_{\pm0.034}$ \\

\multicolumn{1}{c}{}                           & SolvBert       & \multicolumn{1}{c}{$0.602_{\pm0.029}$} & $1.034_{\pm0.044}$ & \multicolumn{1}{c}{{${0.496}_{\pm0.007}$}} & {${0.693}_{\pm0.014}$} & \multicolumn{1}{c}{$0.192_{\pm0.008}$} & $0.353_{\pm0.008}$ & \multicolumn{1}{c}{{${0.418}_{\pm0.018}$}} & {${0.711}_{\pm0.020}$} \\ 

\multicolumn{1}{c}{}                           & Uni-Mol    & \multicolumn{1}{c}{{${0.575}_{\pm0.060}$}} & {${1.012}_{\pm0.070}$} & \multicolumn{1}{c}{$0.355_{\pm0.007}$} & $0.602_{\pm0.024}$ & \multicolumn{1}{c}{$0.198_{\pm0.002}$} & $0.344_{\pm0.003}$ & \multicolumn{1}{c}{$0.267_{\pm0.005}$} & $0.669_{\pm0.017}$ \\

\multicolumn{1}{c}{}                           & SMD & \multicolumn{1}{c}{$0.599_{\pm0.037}$} & $1.202_{\pm0.036}$ & \multicolumn{1}{c}{$0.400_{\pm0.022}$} & $0.646_{\pm0.037}$ & \multicolumn{1}{c}{$0.199_{\pm0.006}$} & $0.348_{\pm0.007}$ & \multicolumn{1}{c}{$0.657_{\pm0.011}$} & $1.023_{\pm0.029}$ \\  \midrule

\multicolumn{1}{c}{\multirow{4}{*}{\thead{LLM \\ Based}}} & Galactica  & \multicolumn{1}{c}{$0.882_{\pm0.010}$} & $1.438_{\pm0.066}$ & \multicolumn{1}{c}{$0.645_{\pm0.008}$} & $1.064_{\pm0.016}$ & \multicolumn{1}{c}{$0.594_{\pm0.006}$} & $0.854_{\pm0.008}$ & \multicolumn{1}{c}{$0.831_{\pm0.018}$} & $1.486_{\pm0.035}$ \\ 

\multicolumn{1}{c}{}                           & Chem T5    & \multicolumn{1}{c}{$0.802_{\pm0.036}$} & $1.377_{\pm0.057}$ & \multicolumn{1}{c}{$0.629_{\pm0.010}$} & $0.910_{\pm0.017}$ & \multicolumn{1}{c}{$0.445_{\pm0.008}$} & $0.734_{\pm0.010}$ & \multicolumn{1}{c}{$0.882_{\pm0.015}$} & $1.297_{\pm0.024}$ \\ 

\multicolumn{1}{c}{}                           & MolCA      & \multicolumn{1}{c}{$0.760_{\pm0.033}$} & $1.271_{\pm0.039}$ & \multicolumn{1}{c}{$0.581_{\pm0.007}$} & $0.897_{\pm0.008}$ & \multicolumn{1}{c}{$0.467_{\pm0.006}$} & $0.716_{\pm0.022}$ & \multicolumn{1}{c}{$0.648_{\pm0.033}$} & $1.125_{\pm0.035}$ \\ 

\multicolumn{1}{c}{}                           & MolT5      & \multicolumn{1}{c}{$0.705_{\pm0.047}$} & $1.135_{\pm0.069}$ & \multicolumn{1}{c}{$0.549_{\pm0.008}$} & $0.832_{\pm0.006}$ & \multicolumn{1}{c}{$0.476_{\pm0.003}$} & $0.695_{\pm0.013}$ & \multicolumn{1}{c}{$0.652_{\pm0.023}$} & $1.124_{\pm0.027}$ \\ \midrule

\multicolumn{2}{c|}{ MolTC (Ours)}     & \multicolumn{1}{c}{\cellcolor{gray!30}$0.502_{\pm0.011}$} & \cellcolor{gray!30}$0.684_{\pm0.042}$ & \multicolumn{1}{c}{\cellcolor{gray!30}$0.194_{\pm0.009}$} & \cellcolor{gray!30}$0.388_{\pm0.010}$ & \multicolumn{1}{c}{\cellcolor{gray!30}$0.171_{\pm0.006}$} & \cellcolor{gray!30}$0.295_{\pm0.004}$ & \multicolumn{1}{c}{\cellcolor{gray!30}$0.172_{\pm0.004}$} & \cellcolor{gray!30}$0.465_{\pm0.008}$ \\ \bottomrule 
\end{tabular}
\end{table*}

\subsection{Experimental Setting}
We evaluate MolTC on twelve well-established downstream molecule interaction tasks involving qualitative and quantitative analysis.
Here we provide a brief overview of our experimental setup. Detailed descriptions 
are presented in the appendix.

\vspace{3pt}
\noindent \textbf{Datasets.} We employ 12 datasets across various domains such as DDI, SSI, and CSI. Specifically,
we collect {\textit{Drugbank}} (Version 5.0.3), {\textit{ZhangDDI}} \cite{zhang2017predicting},  {\textit{ChChMiner}} \cite{zitnik2018stanford},
 {\textit{DeepDDI}} \cite{ryu2018deep},
{\textit{TWOSIDES}} \cite{tatonetti2012data},
 {\textit{Chromophore}} \cite{joung2020experimental},
 {\textit{MNSol}} \cite{marenich2020minnesota},
 {\textit{CompSol}} \cite{moine2017estimation},
 {\textit{Abraham}} \cite{grubbs2010mathematical},
 {\textit{CombiSolv}} \cite{vermeire2021transfer}, \textit{FreeSolv} \cite{mobley2014freesolv} and
 {\textit{CombiSolv-QM}} \cite{vermeire2021transfer}. 

\vspace{3pt}
\noindent \textbf{Baselines.} For a comprehensive evaluation, we conduct various baseline methods encompassing distinct categories such as methods based on: GNNs, DL models other than
GNN, and LLMs. Specifically, For DDI task, we employ \textit{GoGNN} \cite{wang2020gognn}, \textit{MHCADDI} \cite{deac2019drug}, \textit{DeepDDI} \cite{ryu2018deep}, \textit{SSI-DDI}, \textit{CGIB} \cite{CGIB}, \textit{CMRL} \cite{lee2023shift}, \textit{MDF-SA-DDI} \cite{lin2022mdf}, \textit{DSN-DDI} \cite{li2023dsn} as the backbone.
For SSI and CSI tasks, we utilize
\textit{D-MPNN} \cite{vermeire2021transfer}, \textit{SolvBert} \cite{RN479}, \textit{SMD} \cite{RN481}, \textit{CGIB} \cite{CGIB}, \textit{CIGIN} \cite{CIGIN}, \textit{GEM} \cite{fang2022geometry}, \textit{GOVER} \cite{biognn1}, \textit{Uni-Mol} \cite{Uni-mol} as the backbone. Furthermore, all downstream tasks adopt LLM-based methods, such as Galactica \cite{Galactica}, Chem T5 \cite{chemT5}, MolT5 \cite{MolT5} and MolCA \cite{MolCA} as the backbone.

\vspace{3pt}
\noindent \textbf{Metrics.}  
For qualitative tasks, we employ prediction \textit{Accuracy} and \textit{AUC-ROC} (Area Under the Receiver Operating Characteristic curve) as comparative metrics, while for quantitative tasks, \textit{MAE} (Mean Absolute Error) and \textit{RMSE }(Root Mean Square Error) are utilized as the standards.

\begin{table}[t]\footnotesize
\centering
\caption{Performance comparison of various models on different datasets.}
\vspace{-0.6em}
\setlength{\tabcolsep}{1.4pt}
\def\arraystretch{.90}
\label{tab:performance_comparison3}
\begin{tabular}{ccccc}
\toprule
\multicolumn{1}{c}{\multirow{2}{*}{Dataset}} & \multicolumn{1}{c}{\multirow{2}{*}{Metric}} & \multicolumn{1}{c}{\multirow{2}{*}{w/o SMILES}} & \multicolumn{2}{c}{ w/o CoT } \\ \cmidrule(l){4-5}
      &     &    & Broad    &Fine    \\ \midrule  \midrule

\multicolumn{1}{c}{\multirow{4}{*}{DDI}} & Accuracy      & \multicolumn{1}{c}{\cellcolor{gray!30}$6.42_{\pm0.13}$} & $2.01_{\pm0.05}$ & \multicolumn{1}{c}{$-$}\\

&{Rate ($\downarrow$)}      & \multicolumn{1}{c}{\cellcolor{gray!30}$7.08$ \%} & $2.13$ \% & \multicolumn{1}{c}{$-$}\\
\cmidrule(l){2-5}

\multicolumn{1}{c}{} & ACC-AUC      & \multicolumn{1}{c}{\cellcolor{gray!30}$7.87_{\pm0.32}$} & $2.98_{\pm0.08}$ & \multicolumn{1}{c}{$-$}\\

&{Rate ($\downarrow$)}      & \multicolumn{1}{c}{\cellcolor{gray!30}$8.22$ \%} & $3.10$ \%& \multicolumn{1}{c}{$-$}\\
\midrule

\multicolumn{1}{c}{\multirow{4}{*}{SSI}} & MAE      & \multicolumn{1}{c}{$0.025_{\pm0.004}$} & $0.010_{\pm0.002}$ & \multicolumn{1}{c}{\cellcolor{gray!30}$0.036_{\pm0.007}$}\\

&{Rate ($\uparrow$)}      & \multicolumn{1}{c}{$11.32$ \%} & $4.56$ \% & \multicolumn{1}{c}{\cellcolor{gray!30}$16.40$ \%}\\
\cmidrule(l){2-5}

\multicolumn{1}{c}{} & RMSE      &     \multicolumn{1}{c}{$0.045_{\pm0.007}$} & $0.014_{\pm0.003}$ & \multicolumn{1}{c}{\cellcolor{gray!30}$0.054_{\pm0.009}$}\\

&{Rate ($\uparrow$)}      & \multicolumn{1}{c}{$9.47$ \%} & $2.95$ \% & \multicolumn{1}{c}{\cellcolor{gray!30}$11.37$ \%}\\
\midrule

\multicolumn{1}{c}{\multirow{4}{*}{\thead{CSI \\ Abs.}}} & MAE      & \multicolumn{1}{c}{$2.06_{\pm0.11}$} & $0.51_{\pm0.03}$ & \multicolumn{1}{c}{\cellcolor{gray!30}$2.65_{\pm0.16}$}\\

&{Rate ($\uparrow$)}      & \multicolumn{1}{c}{$15.03$ \%} & $3.72$ \% & \multicolumn{1}{c}{\cellcolor{gray!30}$19.34$ \%}\\
\cmidrule(l){2-5}

\multicolumn{1}{c}{} & RMSE      & \multicolumn{1}{c}{$3.37_{\pm0.20}$} & $1.18_{\pm0.12}$ & \multicolumn{1}{c}{\cellcolor{gray!30}$4.84_{\pm0.29}$}\\

&{Rate ($\uparrow$)}      & \multicolumn{1}{c}{$15.18$ \%} & $5.31$ \%& \multicolumn{1}{c}{\cellcolor{gray!30}$21.80$ \%}\\
\midrule

\multicolumn{1}{c}{\multirow{4}{*}{\thead{CSI \\ Emis.}}} & MAE      & \multicolumn{1}{c}{$3.10_{\pm0.17}$} & $0.85_{\pm0.04}$ & \multicolumn{1}{c}{\cellcolor{gray!30}$4.42_{\pm0.36}$}\\

&{Rate ($\uparrow$)}      & \multicolumn{1}{c}{$16.23$ \%} & $5.23$ \% & \multicolumn{1}{c}{\cellcolor{gray!30}$23.14$ \%}\\
\cmidrule(l){2-5}

\multicolumn{1}{c}{} & RMSE      & \multicolumn{1}{c}{$4.99_{\pm0.28}$} & $1.47_{\pm0.12}$ & \multicolumn{1}{c}{\cellcolor{gray!30}$7.29_{\pm0.44}$}\\

&{Rate ($\uparrow$)}      & \multicolumn{1}{c}{$18.34$ \%} & $5.40$ \%& \multicolumn{1}{c}{\cellcolor{gray!30}$26.80$ \%}\\
\midrule

\multicolumn{1}{c}{\multirow{4}{*}{\thead{CSI \\ Life.}}} & MAE      & \multicolumn{1}{c}{\cellcolor{gray!30}$0.085_{\pm0.003}$} & $0.026_{\pm0.002}$ & \multicolumn{1}{c}{$0.072_{\pm0.004}$}\\

&{Rate ($\uparrow$)}      & \multicolumn{1}{c}{\cellcolor{gray!30}$13.70$ \%} & $4.19$ \% & \multicolumn{1}{c}{$11.61$ \%}\\
\cmidrule(l){2-5}

\multicolumn{1}{c}{} & RMSE      & \multicolumn{1}{c}{\cellcolor{gray!30}$0.101_{\pm0.010}$} & $0.034_{\pm0.008}$ & \multicolumn{1}{c}{$0.093_{\pm0.010}$}\\

&{Rate ($\uparrow$)}      & \multicolumn{1}{c}{\cellcolor{gray!30}$12.16$ \%} & $4.09$ \%& \multicolumn{1}{c}{$11.20$ \%}\\
\bottomrule 
\end{tabular}
\end{table}

\subsection{Qualitative Prediction Performance (RQ1)}
Table \ref{tab:performance_comparison} exhibits the performance in qualitative interactive tasks. Due to page width limitations, only a subset of the results is presented, with additional results detailed in the appendix. From Table \ref{tab:performance_comparison}, we deduce the following observations:

\vspace{3pt}
\noindent\textbf{Obs.1:} MolTC consistently outshines its counterparts in qualitative interaction predictions, 
While GNN-based methods demonstrate commendable performance, maintaining over 90\% accuracy across numerous datasets, MolTC transcends these figures in every evaluated scenario. 
For instance, it marks a notable 1.05\% improvement in accuracy on the drugback dataset, a feat attributable to 
the synergy between the LLMs' reasoning faculties and the GNNs' proficiency in graph modeling.

\noindent\textbf{Obs.2:} The variability of MolTC's outcomes, as indicated by the standard deviation, is consistently minimal in comparison to other models,
On average, the standard deviation for MolTC is 35.41\% lower than GNN-based models and 46.86\% lower than LLM-based models. The precision in MolTC's performance is largely 
attributed to the training paradigm enhanced by the multi-hierarchical CoT,
which ensures a meticulous and accurate inference process.

\subsection{Quantitative Prediction Performance (RQ2)} \label{sec:3.3}
Table \ref{tab:performance_comparison2} shows the performance in a subset of quantitative tasks, with an exhaustive set of results detailed in the appendix. The datasets offer four-dimensional molecular information, comprising atom type, chirality tag, bond type, and bond direction. Key observations from Table \ref{tab:performance_comparison2} include:

\vspace{3pt}
\noindent\textbf{Obs.3:} MolTC continues to lead in quantitative analysis tasks, an area typically challenging for LLMs. 
Despite the strong baseline set by CGIB, characterized by low MAE and RMSE across datasets, MolTC outperforms it in every metric. For instance, it achieves a 23.98\% reduction in RMSE on the CombiSolv dataset relative to CGIB. 
This underscores the advantage of adeptly leveraging the interaction between SMILE representations and molecular graph structures.

\noindent\textbf{Obs.4:} LLM-based models, in general, exhibit subpar performance in quantitative tasks compared to traditional DL-based models, attributed to their inadequacy in sharing and transferring learned molecular interaction insights across datasets and the absence of CoT-guided inference.

\subsection{Ablation Study (RQ3)}
Table \ref{tab:performance_comparison3} presents an ablation study aimed at dissecting the influence of SMILE auxiliary analysis and the optimized training paradigms based on Broad-grained and Fine-grained CoT. For the CSI dataset, properties such as the maximum absorption wavelength (Absorption), maximum emission wavelength (Emission), and excited state lifetime (Lifetime) are denoted as Abs., Emis., and Life., respectively. Key observations are as follows:

\vspace{3pt}
\noindent\textbf{Obs.5:} The three studied ablations exhibit significant influence on the results. 
For example, the collective impact of these three ablations registers an average drop of 12.77\%, affirming the substantial enhancement imparted by the proposed strategies.
 
\noindent\textbf{Obs.6:} The most pronounced effect is observed with the ablation of the Fine-grained CoT paradigm, which incurs an average accuracy decrement of 18.82\%. This underscores the pivotal role of guiding the LLM to deduce a numerical range, a strategy particularly beneficial for quantitative analysis tasks, typically a challenging domain for LLMs.

\noindent\textbf{Obs.7:} The least pronounced, yet significant, impact stems from the optimization of the Broad-grained CoT training paradigm, with an average accuracy reduction of 4.35\%. Its importance is particularly underscored for molecular pairs involving larger and more complex molecules, where directly predicting interactive property by LLMs is arduous.

\section{Discussion}
\textbf{Potential Data Leakage Risk.} LLM might have memorized the prediction target during pretraining using a vast amount of text corpus. Hence, we acknowledge that LLM-based methods for biochemistry tasks such as MolT5 \cite{MolT5}, MolCA \cite{MolCA} and our MolTC may face potential data leakage risks due to the memory during the pre-training process of their base LLMs. However, we claim that ablation experiments can mitigate the influence of data leakage, demonstrating the effectiveness of methods despite using potentially contaminated LLMs. For instance, in our ablation experiments, the backbone LLMs are fed with SMILES strings and textual descriptions of molecules solely, and their performance is much worse than that of original methods. This stark contrast leads to two analytical observations:
\begin{itemize}[leftmargin=*]
    \item Firstly, note that these text inputs, highly recognizable and scattered throughout the LLM's training data, could provide all the necessary information to identify a molecule. Hence, if there is significant data leakage, the LLM could easily recall from memory and make reasonably accurate predictions based on these inputs. Based on this, the poor performance of these ablation experiments can, to some extent, counter the concern of data leakage.
    \item Secondly, since these ablation experiments and the original experiments utilized the exact same backbone LLM, the significant differences in their outcomes indicate that the original method's design is highly effective, regardless of whether the backbone LLM is subject to data leakage.
\end{itemize}

In summary, \textbf{while these ablation experiments can not completely eliminate the impact of data leakage, they do mitigate it.} Furthermore, it is worth noting that while the risk of data leakage exists, it does not deter researchers from continuously exploring and optimizing LLM-based frameworks for biochemical tasks. In the future, by constructing new biochemical datasets through biochemical experiments, which do not exist in the LLM's training data, researchers can eliminate the data leakage issue in these frameworks thoroughly.

\section{Conclusion}

This work focuses on molecule rationale learning, which plays a pivotal role in predicting molecular interactions. Specifically, we introduce a novel, unified LLM-based framework for predicting molecular interactive properties, termed MolTC. To efficiently train it, we propose a multi-tiered CoT principle to guide the training paradigm. 
Experiments conducted across twelve varied datasets
demonstrate the superiority of our method over the current GNN
and LLM-based baselines. 
This breakthrough sets a new standard for integrating multimodal data in LLM-based MRL.

\section*{Limitations}
While this research has undergone extensive testing across a diverse array of datasets covering various domains, it does have certain limitations. Specifically, the study has not been subjected to datasets comprising exceptionally large molecules, which represent extreme cases. Furthermore, the methodologies employed in this research have not yet been adapted or evaluated in contexts requiring few-shot or zero-shot learning scenarios.
Future endeavors will focus on expanding the scope of this study to encompass these areas. 

\section*{Ethics Statement}
This work is primarily foundational in molecular relational learning, focusing on the development of a unified LLM-based paradigm. Its primary aim is to contribute to the academic community by enhancing the understanding and implementation of the molecular relational modeling process. We do not foresee any direct, immediate, or negative societal impacts stemming from the outcomes of our research.

\section*{Acknowledgement}
This research is supported by the National Science and Technology Major Project (2023ZD0121102), National Natural Science Foundation of China (92270114).

\bibliography{anthology,custom}
\bibliographystyle{acl_natbib}

\appendix

\section{Related Work}
Since exhaustive experimental validation of the molecule interactions is notoriously
time-consuming and costly \cite{CGIB}, more recently,
adopting LLM has emerged as a promising alternative for efficient and effective molecular relational learning, which are known for their vast knowledge repositories and advanced logical inference capabilities. Compared to the prediction of single-molecule properties \cite{liu2024prott,3dmolm,liu2024simsgt,fang2023knowledge,guo2024what}, the field of molecular interaction prediction is still in its early stages. For instance,

\begin{itemize}[leftmargin=*]
    \item \textbf{Protein-protein interactions (PPI)}. In this context, proteins are represented as residue contact graphs, also known as amino acid graphs, where each node is a residue \cite{LLM4MI1,LLM4MI2,LLM4MI8,LLM4MI8}. Notably, \citet{LLM4MI8} leverages the superior encoding capabilities of the biochemical LLMs, where the input to the LLM is the protein sequence, and the output is a feature vector for each amino acid in the sequence. This output is then used as node features in the residue contact graph to enhance the prediction of PPI tasks.
    \item \textbf{DDI.} In this context, \citet{LLM4MI7} focuses on  enriching cross-modal integration in biology with chemical knowledge and natural language associations, achieving significant results in multiple drug-target interaction prediction tasks. Recently, few-shot drug pair synergy prediction is gradually gaining attention and entering the spotlight\cite{LLM4MI3}.
    \item \textbf{Chemical reactions}. For  understanding chemical reactions \cite{LLM4MI5,guo2024modeling}, \citet{LLM4MI9} selects in-context reaction examples with varying confidence scores closest to the target reaction query, encouraging large models to understand the relationships between these reactions. \citet{LLM4MI4} focuses on optimizing low-sample organic chemical applications by pretraining them with extensive compound libraries and fine-tuning with smaller in-house datasets for specific tasks. \citet{LLM4MI6} introduces a new foundational model, nach0, capable of solving various chemical and biological tasks, including molecular synthesis.

\end{itemize}

More recently, ReactXT \cite{liu2024react} also focuses on modeling molecular interactions using chemical reactions. However, it is unable to leverage CoT for more robust information extraction. Meanwhile, the Mol-instructions dataset has been introduced \cite{fang2023mol} for effectively training biochemistry LLMs, that can be considered as complementary to our MoT-instruction.

\section{Experiments}
Here, we provide a detailed experimental setup along with additional results. It is important to note that for aspects such as dataset division and hyperparameter configurations in baselines, we followed the settings established by CGIB \cite{CGIB}. Moreover, all settings can be found in our code 
\href{https://github.com/MangoKiller/MolTC}{https://github.com/MangoKiller/MolTC}.

\subsection{Datasets}
We employ 12 datasets across various domains such as DDI, SSI and CSI. 

\vspace{3pt}
\noindent\textbf{Drugbank (version 5.0.3).} This dataset consists of 1704 drugs, 191400 drug pairs, and defines 86 distinct DDI event types. Essential drug information, including DrugBank ID, drug name, molecular SMILES, and target.
provided.

\noindent\textbf{ZhangDDI} \cite{zhang2017predicting}. It contains 548 drugs and 48,548 pairwise interaction data and multiple types of similarity information about these drug pairs.

\noindent\textbf{ChChMiner} \cite{zitnik2018stanford}. It contains 1,322 drugs and 48,514 labeled DDIs, obtained through drug labels andscientific publications.

\noindent\textbf{DeepDDI} \cite{ryu2018deep}. It contains 192,284 labeled DDIs and their detailed side-effect information, which is extracted from Drugbank.

\noindent\textbf{TWOSIDES} \cite{tatonetti2012data}. It collects 555 drugs and their 3,576,513 pairwise interactions involving 1318 interaction types from TWOSIDES.

\noindent\textbf{Chromophore} \cite{joung2020experimental}. It contains 20,236 combinations of 7,016 chromophores and 365 solvents which are given in the SMILES
string format. All optical properties are based on scientific publications and unreliable experimental results are excluded after
examination of absorption and emission spectra. In this dataset,
we measure our model performance on predicting maximum
absorption wavelength (Absorption), maximum emission
wavelength (Emission) and excited state lifetime (Lifetime) properties which are important parameters for the design of chromophores for specific applications. We delete the
NaN values to create each dataset which is not reported in the
original scientific publications. Moreover, for Lifetime data, we
use log normalized target value since the target value of the
dataset is highly skewed inducing training instability.
 
\noindent\textbf{MNSol} \cite{marenich2020minnesota}. It contains 3,037 experimental free energies of solvation or transfer energies of 790 unique solutes and 92 solvents.In this work, we consider 2,275 combinations of 372 unique solutes and 86 solvents following previous work.

\noindent\textbf{FreeSolv} \cite{mobley2014freesolv}. It  provides 643 experimental and calculated hydration free energy of small molecules in water. In this work,
we consider 560 experimental results following previous work.

\noindent\textbf{CompSol} \cite{moine2017estimation}. This dataset is proposed to show how solvation energies are influenced by hydrogen-bonding association effects.
We consider 3,548 combinations of 442 unique solutes and 259
solvents in the dataset following previous work.

\noindent\textbf{Abraham} \cite{grubbs2010mathematical}. This  dataset is a collection of data published by the
Abraham research group at College London. We consider 6,091
combinations of 1,038 unique solutes and 122 solvents following
previous work.

\noindent\textbf{CombiSolv} \cite{vermeire2021transfer}. It contains all the data of MNSol, FreeSolv, CompSol, and Abraham, resulting in 10,145 combinations of 1,368
solutes and 291 solvents.

\noindent\textbf{CombiSolv-QM} \cite{vermeire2021transfer}. It is generated
with 1 million combinations of 284 commonly used solvents and 11,029
solutes. Those 1 million data points are randomly selected from all possible solvent–solute combinations. Solvents and solutes with elements
$H, B, C, N, O, F, P, S, Cl, Br$ and $I$ are included with a solute molar
mass ranging from 2.02 g/mol to 1776.89 g/mol.


\begin{table*}[ht]\footnotesize
\centering
\caption{Comparative performance of various methods in qualitative and quantitative
interactive tasks. The best-performing methods are highlighted with a gray background, while the second-best methods are underscored for emphasis.}
\vspace{-0.6em}
\setlength{\tabcolsep}{2.8pt}
\def\arraystretch{1.03}
\label{tab:performance_comparison_app}
\begin{tabular}{c|c|c|cccc|c}
\toprule
\multirow{2}{*}{Domains} & \multirow{2}{*}{Datasets} & \multirow{2}{*}{Metrics} & \multicolumn{4}{c|}{Baselines} & {Ours}\\  
 &  &  & Galactica & Chem T5 & MolCA & MolT5 & MolTC \\ \midrule  \midrule
  \multirow{2}{*}{DDI} &   \multirow{2}{*}{TWOSIDES}   &  ACC         &     $82.01 \pm{1.76}$        &    $84.43 \pm{2.58}$      &    $90.07 \pm{1.86}$          &      $\underline{92.73}\pm{1.65}$      &   \cellcolor{gray!30}$98.42\pm{0.72}$   \\ 

   &     &  AUCROC         &     $87.99\pm{2.41}$   &       $89.52\pm{1.64}$  &      $93.68\pm{0.83}$  &   $\underline{94.00}\pm{0.61}$    &   \cellcolor{gray!30}$99.02\pm{0.14}$    \\   \midrule

     \multirow{2}{*}{SSI} &   \multirow{2}{*}{MNSol}   &  MAE         &     $0.584\pm{0.095}$        &    $0.504\pm{0.038}$      &    $0.491\pm{0.053}$          &      $\underline{0.449}\pm{0.081}$      &   \cellcolor{gray!30}$0.324\pm{0.019}$   \\ 

   &     &  RMSE         &     $1.002\pm{0.101}$   &       $0.973\pm{0.079}$  &      $0.930\pm{0.062}$  &   $\underline{0.858}\pm{0.069}$    &   \cellcolor{gray!30}$0.585\pm{0.023}$    \\   \midrule

        \multirow{3}{*}{CSI} &   {Absorption}   &  RMSE         &     $43.16\pm{1.38}$        &    $38.70\pm{1.84}$      &    $\underline{36.53}\pm{2.03}$          &      $38.01\pm{2.27}$      &   \cellcolor{gray!30}$28.28\pm{2.20}$   \\ 

   &  {Emission}   &  RMSE         &     $49.85\pm{2.47}$   &       $46.18\pm{2.28}$  &      $\underline{43.35}\pm{1.94}$  &   $46.06\pm{1.65}$    &   \cellcolor{gray!30}$35.43\pm{1.88}$    \\  

    &   {Lifetime}  &  RMSE         &     $1.951\pm{0.115}$   &       $1.633\pm{0.069}$  &      $1.480\pm{0.092}$  &   $\underline{1.394}\pm{0.145}$    &   \cellcolor{gray!30}$1.198\pm{0.073}$    \\

\bottomrule
\end{tabular}
\end{table*}

\begin{table}[ht]\footnotesize
\centering
\caption{Comparative performance of various methods in CombiSolv-QM. The best-performing methods are highlighted with a gray background, while the second-best methods are underscored for emphasis.}
\vspace{-0.6em}
\setlength{\tabcolsep}{2.9pt}
\def\arraystretch{1.05}
\label{tab:performance_comparison_app2}
\begin{tabular}{cc|cc}
\toprule
\multicolumn{2}{c|}{\multirow{2}{*}{\thead{Baseline   Model}}}  & \multicolumn{2}{c}{CombiSolv-QM} \\ 
                     &     &MAE           & RMSE               \\ \midrule  \midrule

\multicolumn{1}{c}{\multirow{4}{*}{\thead{GNN \\ Based}}} & CIGIN      & \multicolumn{1}{c}{$0.077_{\pm0.002}$} & $0.176_{\pm0.004}$  \\

\multicolumn{1}{c}{}    & D-MPNN    & \multicolumn{1}{c}{$0.116_{\pm0.006}$} & $0.208_{\pm0.005}$ \\ 

\multicolumn{1}{c}{}      & GEM    & \multicolumn{1}{c}{$0.079_{\pm0.003}$} & $0.162_{\pm0.002}$ \\

\multicolumn{1}{c}{}                           & CGIB       & \multicolumn{1}{c}{$\underline{0.074}_{\pm0.004}$} & $\underline{0.150}_{\pm0.005}$  \\ \midrule

\multicolumn{1}{c}{\multirow{4}{*}{\thead{ML \\ Based}}}                    & GOVER    & \multicolumn{1}{c}{$0.094_{\pm0.003}$} & $0.277_{\pm0.005}$  \\

\multicolumn{1}{c}{}                           & SolvBert       & \multicolumn{1}{c}{$0.102_{\pm0.005}$} & $0.318_{\pm0.006}$ \\ 

\multicolumn{1}{c}{}                           & Uni-Mol    & \multicolumn{1}{c}{{${0.089}_{\pm0.006}$}} & {${0.214}_{\pm0.005}$} \\

\multicolumn{1}{c}{}                           & SMD & \multicolumn{1}{c}{$0.107_{\pm0.004}$} & $0.341_{\pm0.003}$ \\  \midrule

\multicolumn{1}{c}{\multirow{4}{*}{\thead{LLM \\ Based}}} & Galactica  & \multicolumn{1}{c}{$0.303_{\pm0.004}$} & $0.601_{\pm0.008}$ \\ 

\multicolumn{1}{c}{}                           & Chem T5    & \multicolumn{1}{c}{$0.321_{\pm0.006}$} & $0.555_{\pm0.008}$ \\ 

\multicolumn{1}{c}{}                           & MolCA      & \multicolumn{1}{c}{$0.298_{\pm0.004}$} & $0.545_{\pm0.007}$  \\ 

\multicolumn{1}{c}{}                           & MolT5      & \multicolumn{1}{c}{$0.214_{\pm0.004}$} & $0.339_{\pm0.009}$  \\ \midrule

\multicolumn{2}{c|}{ MolTC (Ours)}     & \multicolumn{1}{c}{\cellcolor{gray!30}$0.072_{\pm0.002}$} & \cellcolor{gray!30}$0.140_{\pm0.003}$ \\ \bottomrule 
\end{tabular}
\end{table}

\subsection{Baselines}
We use both specific
task conventional deep learning models and current biochemical LLMs as the baselines. Specifically, for qualitative tasks:

\vspace{5pt}
\noindent\textbf{GoGNN} \cite{wang2020gognn}. It extracts features from structured entity graphs and entity interaction graphs in a hierarchical manner. We also propose a dual attention mechanism that enables the model to preserve the importance of neighbors in both levels of the graph.

\noindent\textbf{MHCADDI} \cite{deac2019drug}. A gated information transfer neural network is used to control the extraction of substructures and then interact based on an attention mechanism.

\noindent\textbf{DeepDDI} \cite{ryu2018deep}. First, the structural similarity profile is calculated between the two input drugs and other drugs, and then prediction is completed based on the deep neural network.

\noindent\textbf{SSI-DDI} \cite{nyamabo2021ssi}. It uses a 4-layer GAT network to extract substructures at different levels, and  completes the final prediction based on the co-attention mechanism.

\noindent\textbf{CGIB} \cite{CGIB}. Based on the graph conditional information bottleneck theory, conditional subgraphs are extracted to complete the interaction between molecules.

\noindent\textbf{CMRL} \cite{lee2023shift}. It detects the core substructure that is causally related to chemical reactions by introducing a novel conditional intervention framework whose intervention is conditioned on the paired molecule. 

\noindent\textbf{MDF-SA-DDI} \cite{lin2022mdf}. It predicts interaction based on multi-source drug fusion, multi-source feature fusion and transformer self-attention mechanism.

\noindent\textbf{DSN-DDI} \cite{li2023dsn}. It employs local and global representation learning modules iteratively and learns drug substructures from the single drug `intra-view') and the drug pair (`inter-view') simultaneously.

\vspace{5pt}
For quantitative task, we employ the following baselines:

\noindent\textbf{D-MPNN} \cite{vermeire2021transfer}. It employs a transfer learning approach to predict solvation free energies, integrating quantum calculation fundamentals with the heightened accuracy of experimental measurements through two new databases, CombiSolv-QM and CombiSolv-Exp.

\noindent\textbf{SolvBert} \cite{RN479}. It interprets solute and solvent interactions through their combined SMILES representation. After pre-training by unsupervised learning with a substantial computational solvation free energy database, SolvBERT is adaptable to predict experimental solvation free energy or solubility by fine-tuning on specific databases.

\noindent\textbf{SMD} \cite{RN481}. It utilizes the quantum charge density of a solute and a continuum representation of the solvent. It breaks down solvation free energy into two components: bulk electrostatic contribution, treated through a self-consistent reaction field using IEF-PCM, and a cavity-dispersion-solvent-structure term, accounting for short-range interactions in the solvation shell based on atomic surface areas with geometry-dependent constants.

\noindent\textbf{CIGIN} \cite{CIGIN}. It  adopts an end-to-end framework consisting of three essential phases: message passing, interaction, and prediction. In the final phase, these stages are leveraged to predict solvation free energies.

\noindent\textbf{GEM} \cite{fang2022geometry}. It exhibits a uniquely designed geometry-based graph neural network architecture, complemented by several dedicated self-supervised learning strategies at the geometry level. That aims to acquire comprehensive molecular geometry knowledge for accurate prediction of molecular properties.

\noindent\textbf{GOVER} \cite{biognn1}. It captures rich structural information from extensive unlabeled molecular data through self-supervised tasks, employing a flexible Transformer-style architecture integrated with Message Passing Networks. This allows GROVER to be trained efficiently on large-scale datasets without supervision, addressing data scarcity and bias challenges.

\noindent\textbf{Uni-Mol} \cite{Uni-mol}. It incorporates two pre-trained models featuring the SE(3) Transformer architecture: a molecular model pre-trained on 209 million molecular conformations and a pocket model pre-trained on 3 million candidate protein pocket data. Additionally, Uni-Mol integrates various fine-tuning strategies to effectively apply these pre-trained models across diverse downstream tasks.

\subsection{Modules} 
In our experiments,  the two graph encoder are instantiated by the five-layer GINE \cite{GINE}. We conduct 2 million molecules from the ZINC15 \cite{ZINC15} dataset to pretrain them by contrastive learning following \cite{MolCA}. Similarly, two projector are initialized with the encoder-only transformer, Sci-BERT, which is pretrained on scientific publications \cite{scibert}, while its cross-attention modules are randomly initialized. More detailed pretraining process of our Q-Formors follows the training process in \cite{MolCA}, such as there are 8 query tokens in Q-Formers ($N_q=8$).  Note that for LLM-based baselines, we fine-tune the backbone LLMs on task-specific datasets for fair comparison. Their prediction is considered accurate only if the outputs include words or numbers that correctly depict the interaction in question, without presenting any that describe alternative interactions. 

\subsection{Training Epochs}
During the fine-tuning phase, the number of epochs varies for different tasks. For example, for the DDI task, we typically fine-tune for 100 epochs. For SSI datasets with more than 3000 molecular pairs, we initially fine-tune on the CombiSolv-QM \cite{vermeire2021transfer} dataset for 100 epochs, followed by an additional 30 epochs on their respective datasets. For SSI datasets with fewer than 3000 molecular pairs, this number is adjusted to 20. Furthermore, both the fine-tuning and pre-training phases employ the same configuration for the optimizer and learning rate scheduler, as detailed in the following section.

\subsection{Training Strategy}
We employ the AdamW optimizer \cite{loshchilov2017decoupled} with a weight decay set at 0.05. Our learning rate strategy utilizes a combination of linear warm-up and cosine decay, optimizing the training process by initially increasing the learning rate to promote faster convergence, and then gradually decreasing it according to a cosine curve to fine-tune the model parameters. 
LoRA is implemented  using the Open Delta library \cite{ding2022delta}, and the PEFT library \cite{peft}. LoRA's rank $r$ is set to 16, while LoRA is applied  to Galactica's modules of \texttt{[q\_proj, v\_proj, out\_proj, fc1, fc2]} following \cite{MolCA}. This configuration yields a LoRA adapter with 12M parameters which constitutes merely 0.94\% of the parametersin the $\text{Galactica}_\text{1.3B}$.

\subsection{More Experimental Results}
Table \ref{tab:performance_comparison_app} presents the experimental results not shown in the main text due to length constraints. Note that the three datasets in the CSI domain are all derived by splitting the Chromophore dataset. As discussed in Section \ref{sec:3.3}, for a fair comparison, we limited the input features to four-dimensional molecular information, comprising atom type, chirality tag, bond type, and bond direction. Given the difficulty of convergence for some DL-based baselines under this setting, we only showcased the performance of the LLM-based baselines. Meanwhile, considering that our SSI tasks are firstly fine-tuned on the CombiSolv-QM dataset, we present the comprehensive results of this dataset, as shown in Table \ref{tab:performance_comparison_app2}. Observations from Table \ref{tab:performance_comparison_app} and \ref{tab:performance_comparison_app2} are largely consistent with those in the main experimental section. That is, across all tasks, our MolTC outperforms the LLM-based baseline methods in a large margin.

\section{Future Work}
In this paper, we introduce a novel unified framework, leveraging LLM technology to predict molecular interactive properties. The future development directions of this project are twofold. First, we aim to adopt advanced graph encoding techniques to enhance the comprehension of molecular graphs \cite{gaoy1,wang2023snowflake,yu2023learning,yu2024text,yuangao_2024_anomaly}. Secondly, in light of the extensive input information introduced in multi-molecular input scenarios, we plan to employ techniques such as graph interpretability \cite{gnnex,PGex,luo_Evaluating_Explainability,fangjf2023exgc} and sparsification \cite{wang2023brave,wang2023searching} to eliminate information redundancy and resist out-of-distribution \cite{OAR,CGE}.



\end{document}